\documentclass[aps,prd,twocolumn]{revtex4-2}
\usepackage[colorlinks=true, linkcolor=blue, citecolor=blue, urlcolor=blue]{hyperref}
\usepackage{bm,latexsym,amssymb,amsmath,mathrsfs}
\usepackage{graphicx}
\usepackage{color}
\usepackage{times,setspace}

\begin{document}

\title{Quantum variational approach to lattice gauge theory at nonzero density}
\author{Arata Yamamoto}
\affiliation{Department of Physics, The University of Tokyo, Tokyo 113-0033, Japan}

\begin{abstract}
The simulation of dense fermionic matters is a long-standing problem in lattice gauge theory.
One hopeful solution would be the use of quantum computers.
In this paper, digital quantum simulation is designed for lattice gauge theory at nonzero density.
The quantum variational algorithm is adopted to obtain the ground state at nonzero density.
A benchmark test is performed in the lattice Schwinger model.
\end{abstract}

\maketitle

\section{Introduction}

Lattice quantum chromodynamics (QCD) at nonzero baryon density suffers from the sign problem.
Many challenging approaches were proposed to overcome the sign problem, but the general solution has not yet been discovered.
The sign problem might be beyond the limit of classical supercomputers.
In recent years, quantum technology is very rapidly developing.
Quantum computers open up new possibilities in lattice QCD \cite{Zohar:2015hwa,Dalmonte:2016alw,Banuls:2019bmf}.
Although the lattice QCD simulation is difficult for the present noisy-intermediate-scale-quantum (NISQ) devices, it might be feasible in future.
Toward the final goal, we should start discussing how to formulate the simulation of dense QCD in quantum computers.

In classical computers, lattice gauge theory is usually formulated by the path integral with the Euclidean action.
This is nothing but thermodynamics with the grand canonical partition function.
Physical observables are calculated as functions of temperature and chemical potential.
On the other hand, in gate-based quantum computers, qubits are state vectors and quantum gates are unitary operations, so the operator formalism with the Hamiltonian is natural.
In the Hamiltonian formalism, it is easier to consider energy eigenstates than thermal average.
When we are interested in the physics at zero temperature, we need only the ground state.
As for density, it is easier to fix particle number density than chemical potential.
Since net electric charge is conserved in gauge theory, we can take the basis where the total number of charged particles is fixed.
Therefore, we can study the physics at zero temperature and nonzero density by two steps: we prepare the ground state with the total particle number fixed, and then calculate physical observables for the prepared state.
This is the standard procedure in condensed matter physics and nuclear physics, where the Hamiltonian formalism is familiar.
Although the Hamiltonian formalism has been considered in the long history of lattice gauge theory \cite{Kogut:1974ag}, we should reconsider it from more concrete point of view.

In this paper, we design the computational strategy and algorithm for the quantum simulation of lattice gauge theory at nonzero density.
We give some demonstrations in a simple theory, the lattice Schwinger model.
For the demonstrations, we used the simulator in IBM Quantum, which is a classical algorithm to simulate quantum computation, not a real quantum computer.
The results are free from any quantum device noise.
When system size is small, the Hamiltonian can be diagonalized and all the eigenstates can be obtained by a classical computer.
We also performed the exact diagonalization for comparison.

\section{Schwinger model}

The lattice Schwinger model is the lattice discretization of quantum electrodynamics in $1+1$ dimensions.
The model is frequently used in the researches of quantum computation, both in theory \cite{Zache:2018jbt,Klco:2018kyo,Magnifico:2019kyj,Chakraborty:2020uhf,Kharzeev:2020kgc,Shaw:2020udc} and experiment \cite{Banerjee:2012pg,Martinez_2016,Muschik:2016tws,Kokail_2019,Surace_2020}.
The theoretical formulation and the device implementation are very well explained in the literature.
We here summarize necessary equations.

The fundamental building blocks are the spinless fermion $\chi_n$, the electric field $L_n$, and the spatial gauge field $U_n$.
The Hamiltonian on the $N$-site lattice is
\begin{eqnarray}
 H &=& H_L + H_\chi
\\
 H_L &=& J \sum_{n=1}^{N-1} L^2_n 
\\
 H_\chi &=& -i w \sum_{n=1}^{N-1} (\chi^\dagger_n U_n \chi_{n+1} - \chi^\dagger_{n+1} U^\dagger_n \chi_{n}) 
,
\end{eqnarray}
where $J$ and $w$ are c-number parameters.
Open boundary condition is assumed.
In the staggered fermion formulation, the fermion $\chi_n$ is assigned as a particle on even lattice sites and as an antiparticle on odd lattice sites.
The fermion number density operator is given by
\begin{equation}
 Q_n = \chi^\dagger_n \chi_n -\frac{1}{2}[1-(-1)^n]
.
\end{equation}
The electric fields and the number density are constrained by the Gauss law
\begin{equation}
\label{eqGL}
 L_n-L_{n-1} = Q_n
.
\end{equation}
The total particle number of the fermions is given by
\begin{equation}
 Q = \sum_{n=1}^{N} Q_n = L_{N} - L_0
.
\end{equation}
We fixed one boundary condition, $L_0=0$, for simplicity.
The other boundary condition is automatic, $L_N=Q$, when the total particle number is given.
For quantum computation, $L_n$ and $U_n$ are eliminated by the Gauss law \eqref{eqGL} and the field redefinition of $\chi_n$, respectively, and $\chi_n$ is transformed to the Pauli matrices by the Jordan-Wigner transformation \cite{Jordan:1928wi}.
Then we obtain
\begin{eqnarray}
 H_L &=& J \sum_{n=1}^{N-1} \left( \sum_{m=1}^n Q_m \right)^2 
\\
 H_\chi &=& \frac{w}{2} \sum_{n=1}^{N-1} (X_nX_{n+1}+Y_nY_{n+1})
\\
 Q_n &=& \frac12 [Z_n+(-1)^n]
.
\end{eqnarray}
Each term of the Hamiltonian is realized by one Pauli gate $\{Z_n\}$ or product of two Pauli gates $\{X_mX_n,Y_mY_n,Z_mZ_n\}$.

The ``nonzero density'' means that the total particle number
\begin{equation}
 q = \langle Q\rangle
\end{equation}
is nonzero.
From the commutation relation $[Q,H]=0$, the total particle number is conserved.
In other words, the Hamiltonian is block-diagonal and the Hilbert space is decomposed into subspaces $q=N/2, \cdots, 1, 0, -1, \cdots, -N/2$.
The states in different particle number sectors cannot mix with each other.
The expectation value $\langle O \rangle$ of a physical observable can be written as a function of $q$ if $O$ does not create or annihilate particles.
Changing the particle number sector by hand, we can study the density-dependence of the physical observable.

\section{Quantum adiabatic algorithm}
\label{sec3}

\begin{figure}[b]
\begin{center}
 \includegraphics[width=.48\textwidth]{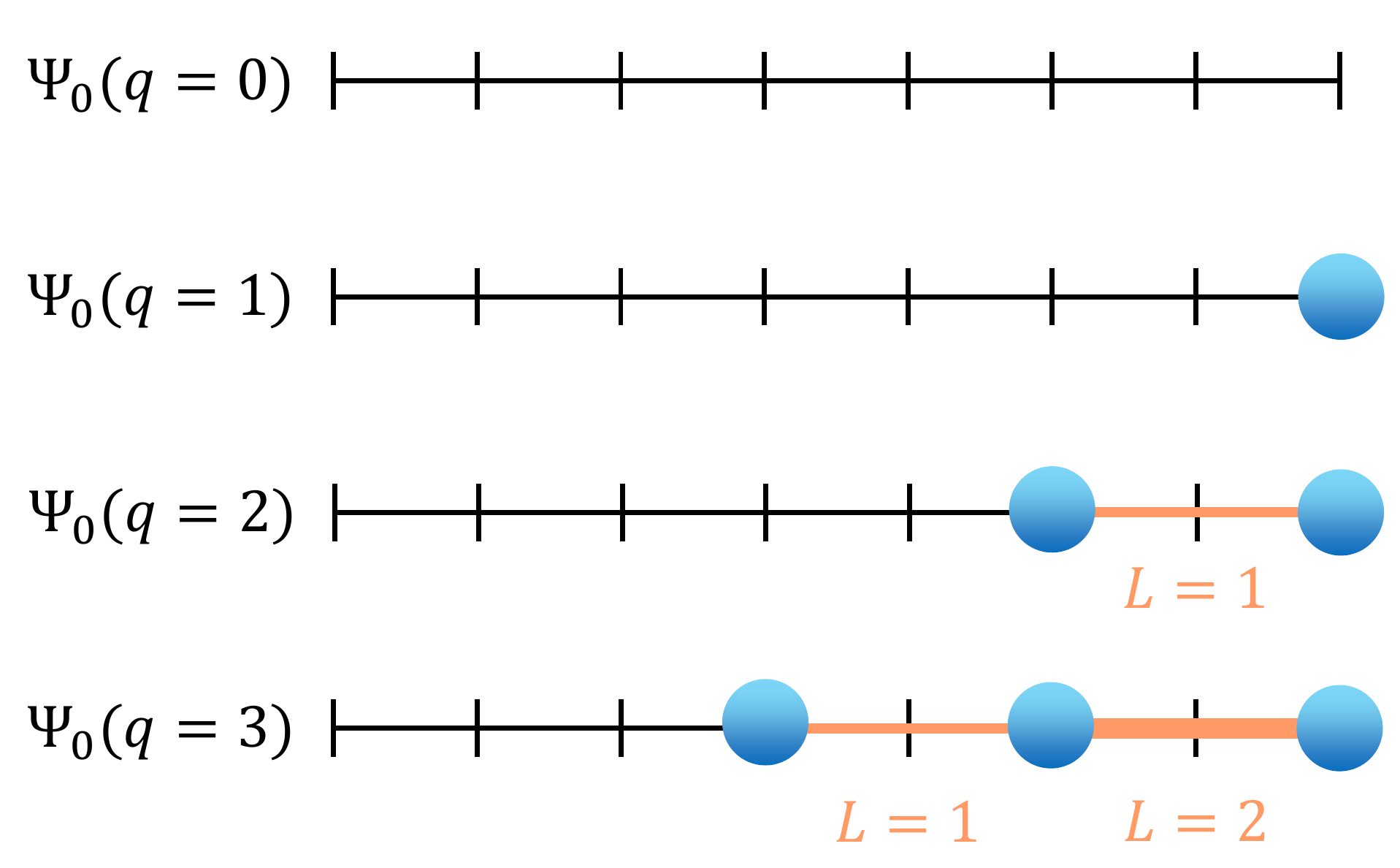}
\caption{
\label{fig}
Ground states $\Psi_0(q)$ of the Hamiltonian $H_L$ on the 8-site lattice.
The electric field at the left boundary is fixed at $L_0=0$.
Once the boundary condition is given, the ground state is easily obtained, e.g., as for $q=3$, three fermions (blue balls) occupy the sites $n=4,6,8$ and induce the electric fields $L_4=L_5=1$ and $L_6=L_7=2$ (orange lines).
}
\end{center}
\end{figure}

The quantum adiabatic algorithm is the computational scheme to obtain the ground state of arbitrary Hamiltonian \cite{Farhi2000,Farhi2001-mi}.
It is based on the adiabatic theorem; if the Hamiltonian is gradually changed from one to another and if the initial state is the ground state, the system keeps staying the ground state.
The quantum adiabatic algorithm was applied to many models, including the lattice Schwinger model with $q=0$ \cite{Chakraborty:2020uhf}.

In the lattice Schwinger model, the ground state of the electric field Hamiltonian $H_L$ is the pure state depicted in Fig.~\ref{fig}.
The initial state $\Psi_0(q)$ is set to the ground state of $H_L$.
The remaining part, the fermion Hamiltonian $H_\chi$, is treated as the perturbation term.
The algorithm is written by
\begin{equation}
 |\Psi(q)\rangle = e^{-i\int_0^T dt \left( H_L+\frac{t}{T}H_\chi \right)} |\Psi_0(q)\rangle
.
\label{eqpsia}
\end{equation}
In the limit of $T\to \infty$, the evolution is adiabatic and the ground state of $H$ is obtained.
The evolution operator is approximated by the Suzuki-Trotter decomposition,
\begin{eqnarray}
&&  e^{-i\int_0^T dt \left( H_L+\frac{t}{T}H_\chi \right)} \simeq U(S) U(S-1)\cdots U(1)
\\
&& U(s) = e^{-i\frac{\delta t}{2} \frac{s}{S} H_\chi} e^{-i\delta t H_L} e^{-i\frac{\delta t}{2} \frac{s}{S} H_\chi} 
\end{eqnarray}
with $t= s \delta t$ and $T=S \delta t$, where $s$ and $S$ are integers.
The integer $S$ must be large enough and the step size $\delta t$ must be small enough that the calculation reproduces the adiabatic theorem.
In quantum circuits, each of $U(s)$ is further decomposed into rotation gates,
\begin{equation}
\begin{split}
 U(s) \simeq& \left[ \prod_n e^{-i\frac{\delta t}{2} \frac{s}{S} \frac{w}{2}Y_nY_{n+1}} e^{-i\frac{\delta t}{2} \frac{s}{S} \frac{w}{2}X_nX_{n+1}} \right]
\\
&\times \left[ \prod_n e^{-i\delta tJ \left( \sum_m Q_m \right)^2} \right]
\\
&\times \left[ \prod_n e^{-i\frac{\delta t}{2} \frac{s}{S} \frac{w}{2}X_nX_{n+1}} e^{-i\frac{\delta t}{2} \frac{s}{S} \frac{w}{2}Y_nY_{n+1}} \right]
.
\label{equ}
\end{split}
\end{equation}
We can show the commutation relation $[Q,U(s)]=0$.
The commutation relation guarantees that the obtained state $\Psi(q)$ has the same total particle number as the initial state $\Psi_0(q)$.
There is a technical but important issue.
In general, the decomposition is not unique.
For example, another choice is $U(s)= \left[\prod_n e^{-i\frac{\delta t}{2} \frac{s}{S} \frac{w}{2}Y_nY_{n+1}} \right] \left[ \prod_n e^{-i\frac{\delta t}{2} \frac{s}{S} \frac{w}{2}X_nX_{n+1}} \right]\cdots$.
This choice is however bad because it does not satisfy the commutation relation.
We must choose the decomposition which conserves the total particle number, otherwise the obtained state $\Psi(q)$ is not the eigenstate of $Q$.

\begin{figure}[t]
\begin{center}
 \includegraphics[width=.48\textwidth]{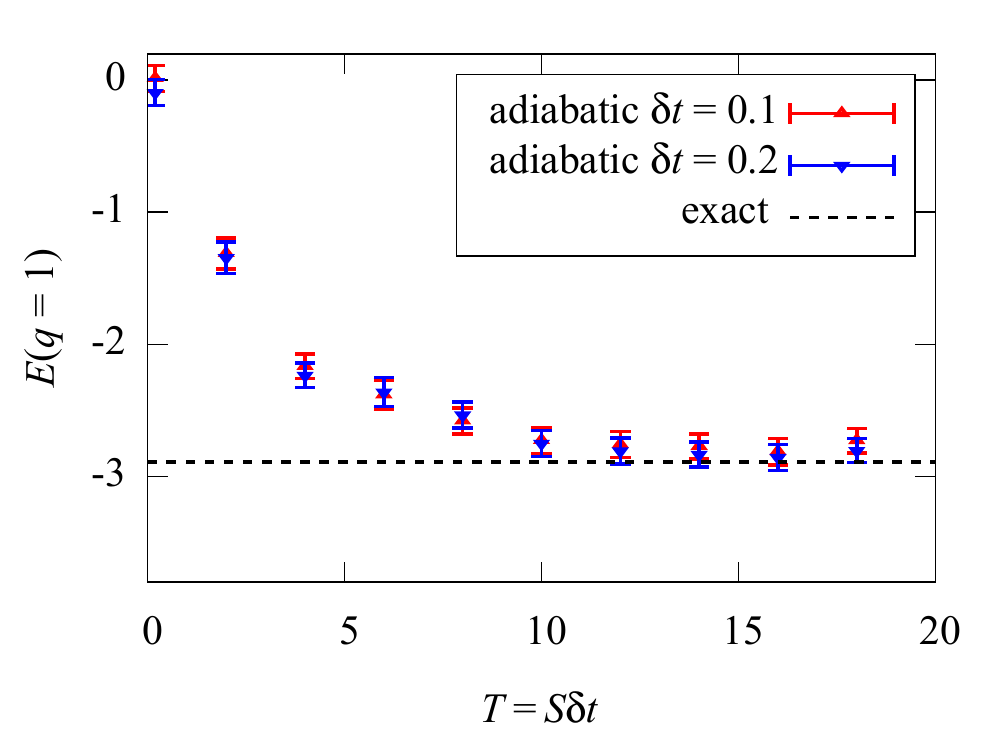}
\caption{
\label{figEA}
Ground-state energy $E(q=1)$ computed by the quantum adiabatic algorithm.
The data with the step sizes $\delta t = 0.1$ and $0.2$ are shown.
The broken line is the exact value obtained by matrix diagonalization.
}
\end{center}
\end{figure}

We demonstrate the quantum adiabatic algorithm in the lattice Schwinger model at $q\neq 0$.
We fix the parameters $J=w=1$ and use the dimensionless unit scaled by them.
The lattice size is $N=8$, so the dimension of the Hilbert space is $2^N=256$.
The ground-state energy
\begin{equation}
 E(q) = \langle\Psi(q)|H|\Psi(q)\rangle
\end{equation}
is shown in Fig.~\ref{figEA}.
The energy decreases as $T$ increases, and eventually in $T>10$, becomes insensitive to $T$ (and $\delta t$).
This is exactly the expected behavior in this algorithm.
The energy in $T>10$ is indeed consistent with the exact ground-state energy obtained by matrix diagonalization.
The calculation seems successful, but it has a potential problem.
The operation $U(s)$ needs to be manipulated many times; e.g., $S=50$ for $\delta t =0.2$ and $S=100$ for $\delta t =0.1$ in Fig.~\ref{figEA}.
In real NISQ devices, that enhances decoherence and makes quantum error out of control.
A noise-robust algorithm is required for the real quantum simulation.

\section{Quantum variational algorithm}

We adopt the quantum variational algorithm to suppress the noise.
The prototype of the quantum variational algorithm was proposed in quantum chemistry \cite{Peruzzo2014-hk}, and then applied to other fields, in particular, to the lattice Schwinger model \cite{Klco:2018kyo,Kokail_2019}.
The algorithm consists of the following quantum and classical processes:
(a) One generates a certain quantum state and calculate its energy in a quantum computer.
(b) Based on the calculated energy, one classically determines the feedback to improve the quantum state.
The processes (a) and (b) are repeated until the energy converges to the minimum.

The quality of the variational calculation strongly depends on the variational ansatz, i.e., how to parameterize quantum states.
We take the variational ansatz \cite{farhi2014quantum,Wecker_2015}
\begin{equation}
 |\Psi(q)\rangle = V(S)V(S-1)\cdots V(1) |\Psi_0(q)\rangle
\label{eqpsiv}
\end{equation}
with
\begin{equation}
\begin{split}
 V(s) =& \left[ \prod_n e^{-i\frac{\beta_s}{2}\frac{w}{2}Y_nY_{n+1}} e^{-i\frac{\beta_s}{2}\frac{w}{2}X_nX_{n+1}} \right]
\\
&\times \left[ \prod_n e^{-i\gamma_s J\left( \sum_m Q_m \right)^2} \right]
\\
&\times \left[ \prod_n e^{-i\frac{\beta_s}{2}\frac{w}{2}X_nX_{n+1}} e^{-i\frac{\beta_s}{2}\frac{w}{2}Y_nY_{n+1}} \right]
.
\label{eqv}
\end{split}
\end{equation}
There are $2S$ variational parameters $\{\gamma_s,\beta_s\}(s=1,\cdots,S)$.
Because $V(s)$ is the same form as $U(s)$ in Eq.~\eqref{equ}, it satisfies the commutation relation $[Q,V(s)]=0$ and conserves the total particle number.
It is trivial that the results are improved by increasing the integer $S$ because the number of variational parameters increases.
In this work, we fix the number of variational parameters by further simplification.
Let us consider two parameterizations: the one-parameter ($\alpha$) ansatz
\begin{equation}
\gamma_s = \alpha,
\quad
\beta_s = \alpha \frac{s}{S}
\label{eqbeta1}
\end{equation}
and the two-parameter ($\alpha,\beta$) ansatz
\begin{equation}
\gamma_s = \alpha,
\quad
 \beta_s=\beta \frac{s}{S}
.
\label{eqbeta2}
\end{equation}
They are inspired by the quantum adiabatic algorithm.
In the limit of $S \to \infty$, they can reproduce Eqs.~\eqref{eqpsia} to \eqref{equ}, so the convergence to the exact ground state is a priori ensured.

\begin{figure}[t]
\begin{center}
 \includegraphics[width=.48\textwidth]{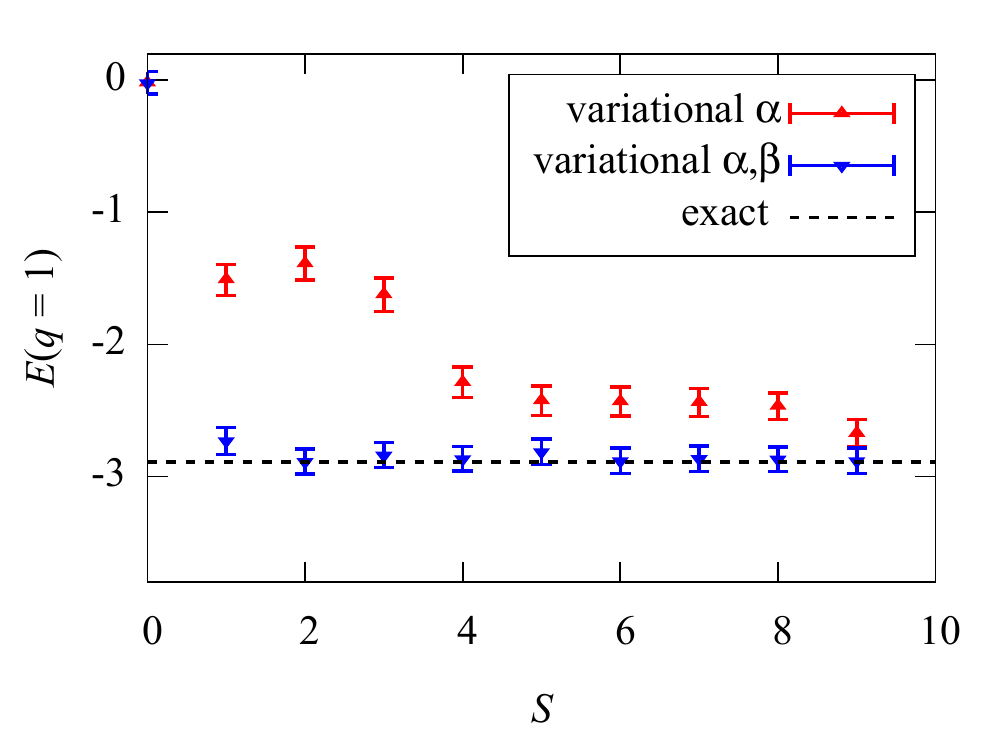}
\caption{
\label{figEV}
Ground-state energy $E(q=1)$ computed by the quantum variational algorithm.
The data with the variational ansatzes \eqref{eqbeta1} and \eqref{eqbeta2} are shown.
The broken line is the exact value obtained by matrix diagonalization.
}
\end{center}
\end{figure}

\begin{figure}[h]
\begin{center}
 \includegraphics[width=.48\textwidth]{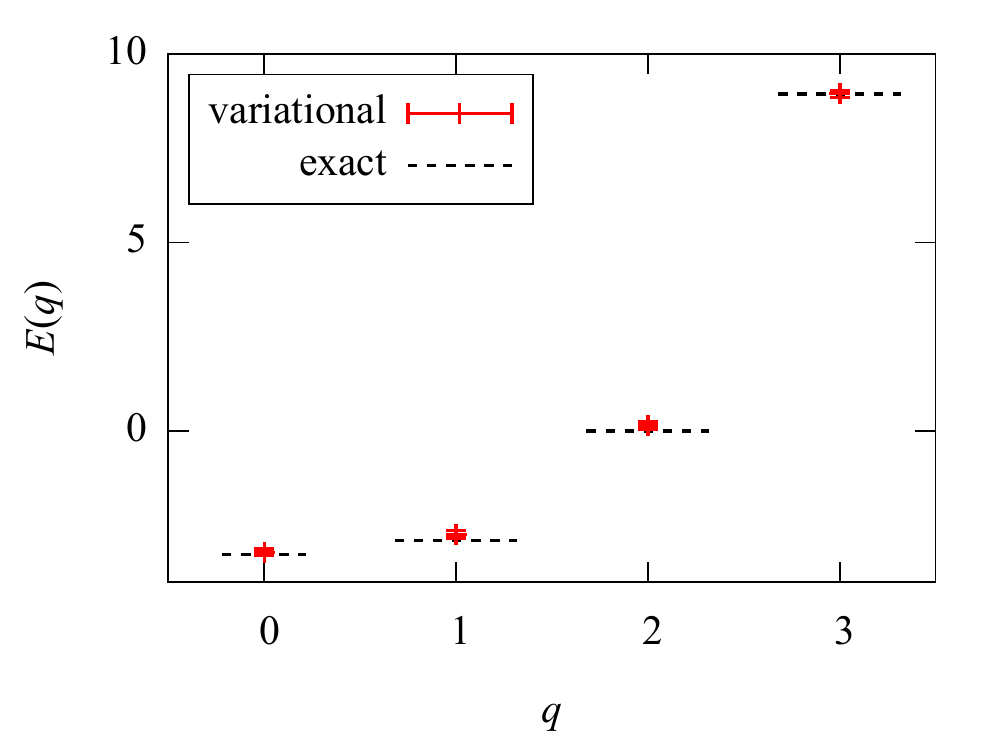}
\caption{
\label{figEQ}
Ground-state energy $E(q)$ as a function of the total particle number $q$.
The data are obtained by the quantum variational algorithm with the ansatz \eqref{eqbeta2} and $S=2$.
The broken lines are the exact values obtained by matrix diagonalization.
}
\end{center}
\end{figure}

We tested the variational calculation with the same simulation condition as in the adiabatic calculation in Sec.~\ref{sec3}.
The dependence of the ground-state energy on the integer $S$ is shown in Fig.~\ref{figEV}.
While the one-parameter ansatz gradually converges as a function of $S$, the two-parameter ansatz rapidly converges and reproduces the exact value even at $S=2$.
The two-parameter ansatz also works well in other particle number sectors, as shown in Fig.~\ref{figEQ}.
The value of $S$ is much smaller than that of the adiabatic calculation in Sec.~\ref{sec3}.
The calculation will be less influenced by the device noise in real quantum computers.

\begin{figure}[t]
\begin{center}
 \includegraphics[width=.48\textwidth]{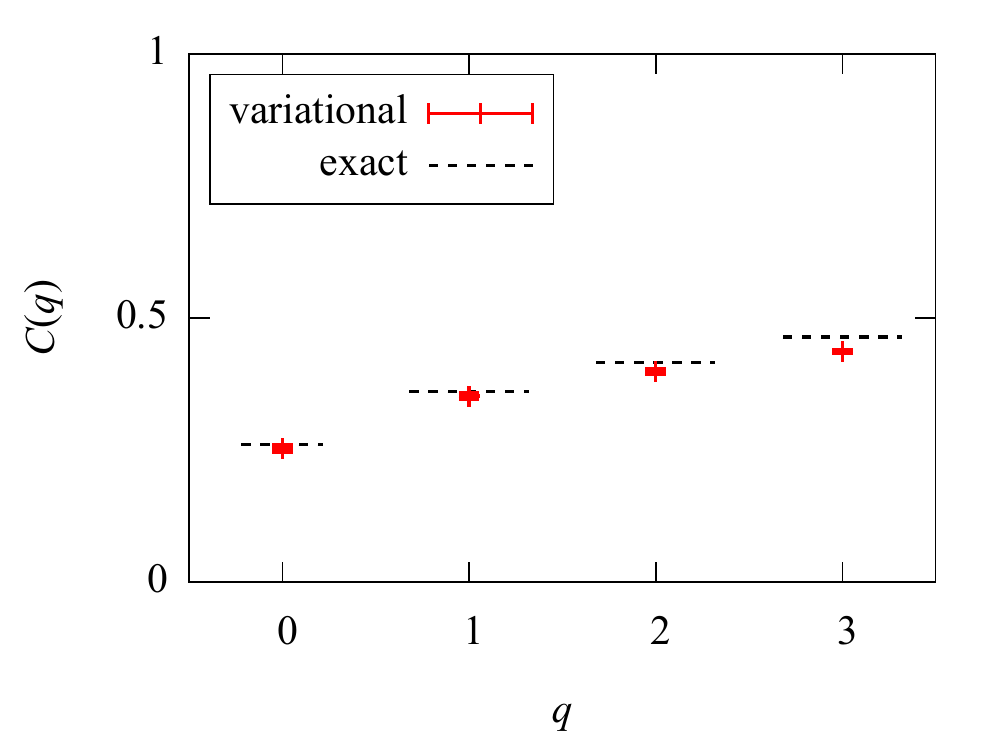}
\caption{
\label{figCQ}
Same as Fig.~\ref{figEQ} but the chiral condensate $C$.
}
\end{center}
\end{figure}

In the analysis of nonzero density, we often want to know the density-dependence of physical observables.
This is now possible because we obtained the state vector of the ground state in each particle number sector.
As an example, we calculated the volume-averaged chiral condensate
\begin{equation}
 C(q) = \frac{1}{N} \sum_{n=1}^N \langle\Psi(q)| (-1)^n \chi^\dagger_n \chi_n |\Psi(q)\rangle + \frac{1}{2}
.
\end{equation}
The second term is the offset to make the chiral condensate in the perturbative vacuum, $\Psi_0(q=0)$ in Fig.~\ref{fig}, zero.
The chiral condensate is plotted as a function of the particle number in Fig.~\ref{figCQ}.
We see that the chiral condensate matches the exact value with good accuracy.
Note that the purpose of this calculation is just to show that we can calculate any observable other than energy.
Physically speaking, the chiral condensate in the massless lattice Schwinger model is lattice artifact.
(The particle-antiparticle symmetry is realized by the even-odd site symmetry in the staggered fermion formulation, but it is explicitly broken by the boundary condition.
This explicit symmetry breaking induces the nonzero chiral condensate.)
We can do physical analysis if we use physical theory.

The results almost perfectly agree with the exact values in this small system.
In realistic theory with large Hilbert space, such perfect agreement is not expected.
We can only obtain an approximate ground state.
To improve the quality of the approximation, we need to increase $S$ or the number of variational parameters.
The decoherence will become non-negligible if $S$ is large and the computational cost to search the minimum will blow up if the number of variational parameters is large; this is the trade-off between the quantum noise and the classical computational cost.

\section{Toward dense QCD}

In this work, we proposed the possibility of quantum computers as a tool to study lattice gauge theory at nonzero density.
The lattice Schwinger model was used because it is the best we can simulate with the current resource of quantum computers.
The basic strategy is quite general and applicable to other lattice gauge theories, e.g., in two- or three-dimensions or with non-Abelian gauge group.
We might be able to challenge dense QCD in a far future.

This approach has two differences from the conventional lattice gauge theory.
One is the computational device: quantum computers or classical computers.
Since this approach does not rely on the Monte Carlo method, it does not encounter the sign problem.
(The data obtained by quantum computers have statistical error, but its origin is essentially different.)
In general, deterministic calculation has the problem that computational time exponentially increases as system size increases, instead of the sign problem.
The variational calculation must have the same problem if it is executed in classical computers.
Even though quantum computation speeds up one part of the calculation, it is non-trivial whether the total computational cost is reasonable or not.
We need further investigation to estimate it.
The other is the theoretical formalism: the Hamiltonian with fixed particle number or the Euclidean path integral with fixed chemical potential.
Almost all the physics can be discussed in equivalent manners, but there are a few exceptions.
An apparent difference is whether the axis of the phase diagram is number density or chemical potential.
Along the number density axis, we cannot meet the phenomenon below the critical chemical potential, such as the Silver Blaze phenomenon.
Finally, we should comment on another proposal of lattice gauge theory at nonzero density \cite{Clemente:2020lpr}, which is based on the Euclidean path integral.
We are not ready to say which is better at the moment, so we should try both possibilities.

\begin{acknowledgments}
The author was supported by JSPS KAKENHI Grant No.~19K03841.   
The author acknowledges the use of IBM Quantum services for this work.
The views expressed are those of the author, and do not reflect the official policy or position of IBM or the IBM Quantum team.
\end{acknowledgments}

\bibliographystyle{apsrev4-2}
\bibliography{paper}

\begin{thebibliography}{22}%
\makeatletter
\providecommand \@ifxundefined [1]{%
 \@ifx{#1\undefined}
}%
\providecommand \@ifnum [1]{%
 \ifnum #1\expandafter \@firstoftwo
 \else \expandafter \@secondoftwo
 \fi
}%
\providecommand \@ifx [1]{%
 \ifx #1\expandafter \@firstoftwo
 \else \expandafter \@secondoftwo
 \fi
}%
\providecommand \natexlab [1]{#1}%
\providecommand \enquote  [1]{``#1''}%
\providecommand \bibnamefont  [1]{#1}%
\providecommand \bibfnamefont [1]{#1}%
\providecommand \citenamefont [1]{#1}%
\providecommand \href@noop [0]{\@secondoftwo}%
\providecommand \href [0]{\begingroup \@sanitize@url \@href}%
\providecommand \@href[1]{\@@startlink{#1}\@@href}%
\providecommand \@@href[1]{\endgroup#1\@@endlink}%
\providecommand \@sanitize@url [0]{\catcode `\\12\catcode `\$12\catcode
  `\&12\catcode `\#12\catcode `\^12\catcode `\_12\catcode `\%12\relax}%
\providecommand \@@startlink[1]{}%
\providecommand \@@endlink[0]{}%
\providecommand \url  [0]{\begingroup\@sanitize@url \@url }%
\providecommand \@url [1]{\endgroup\@href {#1}{\urlprefix }}%
\providecommand \urlprefix  [0]{URL }%
\providecommand \Eprint [0]{\href }%
\providecommand \doibase [0]{https://doi.org/}%
\providecommand \selectlanguage [0]{\@gobble}%
\providecommand \bibinfo  [0]{\@secondoftwo}%
\providecommand \bibfield  [0]{\@secondoftwo}%
\providecommand \translation [1]{[#1]}%
\providecommand \BibitemOpen [0]{}%
\providecommand \bibitemStop [0]{}%
\providecommand \bibitemNoStop [0]{.\EOS\space}%
\providecommand \EOS [0]{\spacefactor3000\relax}%
\providecommand \BibitemShut  [1]{\csname bibitem#1\endcsname}%
\let\auto@bib@innerbib\@empty
%</preamble>
\bibitem [{\citenamefont {Zohar}\ \emph {et~al.}(2016)\citenamefont {Zohar},
  \citenamefont {Cirac},\ and\ \citenamefont {Reznik}}]{Zohar:2015hwa}%
  \BibitemOpen
  \bibfield  {author} {\bibinfo {author} {\bibfnamefont {E.}~\bibnamefont
  {Zohar}}, \bibinfo {author} {\bibfnamefont {J.~I.}\ \bibnamefont {Cirac}},\
  and\ \bibinfo {author} {\bibfnamefont {B.}~\bibnamefont {Reznik}},\ }\href
  {https://doi.org/10.1088/0034-4885/79/1/014401} {\bibfield  {journal}
  {\bibinfo  {journal} {Rept. Prog. Phys.}\ }\textbf {\bibinfo {volume} {79}},\
  \bibinfo {pages} {014401} (\bibinfo {year} {2016})},\ \Eprint
  {https://arxiv.org/abs/1503.02312} {arXiv:1503.02312 [quant-ph]} \BibitemShut
  {NoStop}%
\bibitem [{\citenamefont {Dalmonte}\ and\ \citenamefont
  {Montangero}(2016)}]{Dalmonte:2016alw}%
  \BibitemOpen
  \bibfield  {author} {\bibinfo {author} {\bibfnamefont {M.}~\bibnamefont
  {Dalmonte}}\ and\ \bibinfo {author} {\bibfnamefont {S.}~\bibnamefont
  {Montangero}},\ }\href {https://doi.org/10.1080/00107514.2016.1151199}
  {\bibfield  {journal} {\bibinfo  {journal} {Contemp. Phys.}\ }\textbf
  {\bibinfo {volume} {57}},\ \bibinfo {pages} {388} (\bibinfo {year} {2016})},\
  \Eprint {https://arxiv.org/abs/1602.03776} {arXiv:1602.03776
  [cond-mat.quant-gas]} \BibitemShut {NoStop}%
\bibitem [{\citenamefont {Ba\~nuls}\ \emph {et~al.}(2020)\citenamefont
  {Ba\~nuls} \emph {et~al.}}]{Banuls:2019bmf}%
  \BibitemOpen
  \bibfield  {author} {\bibinfo {author} {\bibfnamefont {M.~C.}\ \bibnamefont
  {Ba\~nuls}} \emph {et~al.},\ }\href
  {https://doi.org/10.1140/epjd/e2020-100571-8} {\bibfield  {journal} {\bibinfo
   {journal} {Eur. Phys. J. D}\ }\textbf {\bibinfo {volume} {74}},\ \bibinfo
  {pages} {165} (\bibinfo {year} {2020})},\ \Eprint
  {https://arxiv.org/abs/1911.00003} {arXiv:1911.00003 [quant-ph]} \BibitemShut
  {NoStop}%
\bibitem [{\citenamefont {Kogut}\ and\ \citenamefont
  {Susskind}(1975)}]{Kogut:1974ag}%
  \BibitemOpen
  \bibfield  {author} {\bibinfo {author} {\bibfnamefont {J.~B.}\ \bibnamefont
  {Kogut}}\ and\ \bibinfo {author} {\bibfnamefont {L.}~\bibnamefont
  {Susskind}},\ }\href {https://doi.org/10.1103/PhysRevD.11.395} {\bibfield
  {journal} {\bibinfo  {journal} {Phys. Rev. D}\ }\textbf {\bibinfo {volume}
  {11}},\ \bibinfo {pages} {395} (\bibinfo {year} {1975})}\BibitemShut
  {NoStop}%
\bibitem [{\citenamefont {Zache}\ \emph {et~al.}(2018)\citenamefont {Zache},
  \citenamefont {Hebenstreit}, \citenamefont {Jendrzejewski}, \citenamefont
  {Oberthaler}, \citenamefont {Berges},\ and\ \citenamefont
  {Hauke}}]{Zache:2018jbt}%
  \BibitemOpen
  \bibfield  {author} {\bibinfo {author} {\bibfnamefont {T.~V.}\ \bibnamefont
  {Zache}}, \bibinfo {author} {\bibfnamefont {F.}~\bibnamefont {Hebenstreit}},
  \bibinfo {author} {\bibfnamefont {F.}~\bibnamefont {Jendrzejewski}}, \bibinfo
  {author} {\bibfnamefont {M.~K.}\ \bibnamefont {Oberthaler}}, \bibinfo
  {author} {\bibfnamefont {J.}~\bibnamefont {Berges}},\ and\ \bibinfo {author}
  {\bibfnamefont {P.}~\bibnamefont {Hauke}},\ }\href
  {https://doi.org/10.1088/2058-9565/aac33b} {\bibfield  {journal} {\bibinfo
  {journal} {Quantum Sci. Technol.}\ }\textbf {\bibinfo {volume} {3}},\
  \bibinfo {pages} {034010} (\bibinfo {year} {2018})},\ \Eprint
  {https://arxiv.org/abs/1802.06704} {arXiv:1802.06704 [cond-mat.quant-gas]}
  \BibitemShut {NoStop}%
\bibitem [{\citenamefont {Klco}\ \emph {et~al.}(2018)\citenamefont {Klco},
  \citenamefont {Dumitrescu}, \citenamefont {McCaskey}, \citenamefont {Morris},
  \citenamefont {Pooser}, \citenamefont {Sanz}, \citenamefont {Solano},
  \citenamefont {Lougovski},\ and\ \citenamefont {Savage}}]{Klco:2018kyo}%
  \BibitemOpen
  \bibfield  {author} {\bibinfo {author} {\bibfnamefont {N.}~\bibnamefont
  {Klco}}, \bibinfo {author} {\bibfnamefont {E.~F.}\ \bibnamefont
  {Dumitrescu}}, \bibinfo {author} {\bibfnamefont {A.~J.}\ \bibnamefont
  {McCaskey}}, \bibinfo {author} {\bibfnamefont {T.~D.}\ \bibnamefont
  {Morris}}, \bibinfo {author} {\bibfnamefont {R.~C.}\ \bibnamefont {Pooser}},
  \bibinfo {author} {\bibfnamefont {M.}~\bibnamefont {Sanz}}, \bibinfo {author}
  {\bibfnamefont {E.}~\bibnamefont {Solano}}, \bibinfo {author} {\bibfnamefont
  {P.}~\bibnamefont {Lougovski}},\ and\ \bibinfo {author} {\bibfnamefont
  {M.}~\bibnamefont {Savage}},\ }\href
  {https://doi.org/10.1103/PhysRevA.98.032331} {\bibfield  {journal} {\bibinfo
  {journal} {Phys. Rev. A}\ }\textbf {\bibinfo {volume} {98}},\ \bibinfo
  {pages} {032331} (\bibinfo {year} {2018})},\ \Eprint
  {https://arxiv.org/abs/1803.03326} {arXiv:1803.03326 [quant-ph]} \BibitemShut
  {NoStop}%
\bibitem [{\citenamefont {Magnifico}\ \emph {et~al.}(2020)\citenamefont
  {Magnifico}, \citenamefont {Dalmonte}, \citenamefont {Facchi}, \citenamefont
  {Pascazio}, \citenamefont {Pepe},\ and\ \citenamefont
  {Ercolessi}}]{Magnifico:2019kyj}%
  \BibitemOpen
  \bibfield  {author} {\bibinfo {author} {\bibfnamefont {G.}~\bibnamefont
  {Magnifico}}, \bibinfo {author} {\bibfnamefont {M.}~\bibnamefont {Dalmonte}},
  \bibinfo {author} {\bibfnamefont {P.}~\bibnamefont {Facchi}}, \bibinfo
  {author} {\bibfnamefont {S.}~\bibnamefont {Pascazio}}, \bibinfo {author}
  {\bibfnamefont {F.~V.}\ \bibnamefont {Pepe}},\ and\ \bibinfo {author}
  {\bibfnamefont {E.}~\bibnamefont {Ercolessi}},\ }\href
  {https://doi.org/10.22331/q-2020-06-15-281} {\bibfield  {journal} {\bibinfo
  {journal} {Quantum}\ }\textbf {\bibinfo {volume} {4}},\ \bibinfo {pages}
  {281} (\bibinfo {year} {2020})},\ \Eprint {https://arxiv.org/abs/1909.04821}
  {arXiv:1909.04821 [quant-ph]} \BibitemShut {NoStop}%
\bibitem [{\citenamefont {Chakraborty}\ \emph {et~al.}(2020)\citenamefont
  {Chakraborty}, \citenamefont {Honda}, \citenamefont {Izubuchi}, \citenamefont
  {Kikuchi},\ and\ \citenamefont {Tomiya}}]{Chakraborty:2020uhf}%
  \BibitemOpen
  \bibfield  {author} {\bibinfo {author} {\bibfnamefont {B.}~\bibnamefont
  {Chakraborty}}, \bibinfo {author} {\bibfnamefont {M.}~\bibnamefont {Honda}},
  \bibinfo {author} {\bibfnamefont {T.}~\bibnamefont {Izubuchi}}, \bibinfo
  {author} {\bibfnamefont {Y.}~\bibnamefont {Kikuchi}},\ and\ \bibinfo {author}
  {\bibfnamefont {A.}~\bibnamefont {Tomiya}},\ }\href@noop {} {\  (\bibinfo
  {year} {2020})},\ \Eprint {https://arxiv.org/abs/2001.00485}
  {arXiv:2001.00485 [hep-lat]} \BibitemShut {NoStop}%
\bibitem [{\citenamefont {Kharzeev}\ and\ \citenamefont
  {Kikuchi}(2020)}]{Kharzeev:2020kgc}%
  \BibitemOpen
  \bibfield  {author} {\bibinfo {author} {\bibfnamefont {D.~E.}\ \bibnamefont
  {Kharzeev}}\ and\ \bibinfo {author} {\bibfnamefont {Y.}~\bibnamefont
  {Kikuchi}},\ }\href {https://doi.org/10.1103/PhysRevResearch.2.023342}
  {\bibfield  {journal} {\bibinfo  {journal} {Phys. Rev. Res.}\ }\textbf
  {\bibinfo {volume} {2}},\ \bibinfo {pages} {023342} (\bibinfo {year}
  {2020})},\ \Eprint {https://arxiv.org/abs/2001.00698} {arXiv:2001.00698
  [hep-ph]} \BibitemShut {NoStop}%
\bibitem [{\citenamefont {Shaw}\ \emph {et~al.}(2020)\citenamefont {Shaw},
  \citenamefont {Lougovski}, \citenamefont {Stryker},\ and\ \citenamefont
  {Wiebe}}]{Shaw:2020udc}%
  \BibitemOpen
  \bibfield  {author} {\bibinfo {author} {\bibfnamefont {A.~F.}\ \bibnamefont
  {Shaw}}, \bibinfo {author} {\bibfnamefont {P.}~\bibnamefont {Lougovski}},
  \bibinfo {author} {\bibfnamefont {J.~R.}\ \bibnamefont {Stryker}},\ and\
  \bibinfo {author} {\bibfnamefont {N.}~\bibnamefont {Wiebe}},\ }\href
  {https://doi.org/10.22331/q-2020-08-10-306} {\bibfield  {journal} {\bibinfo
  {journal} {Quantum}\ }\textbf {\bibinfo {volume} {4}},\ \bibinfo {pages}
  {306} (\bibinfo {year} {2020})},\ \Eprint {https://arxiv.org/abs/2002.11146}
  {arXiv:2002.11146 [quant-ph]} \BibitemShut {NoStop}%
\bibitem [{\citenamefont {Banerjee}\ \emph {et~al.}(2012)\citenamefont
  {Banerjee}, \citenamefont {Dalmonte}, \citenamefont {Muller}, \citenamefont
  {Rico}, \citenamefont {Stebler}, \citenamefont {Wiese},\ and\ \citenamefont
  {Zoller}}]{Banerjee:2012pg}%
  \BibitemOpen
  \bibfield  {author} {\bibinfo {author} {\bibfnamefont {D.}~\bibnamefont
  {Banerjee}}, \bibinfo {author} {\bibfnamefont {M.}~\bibnamefont {Dalmonte}},
  \bibinfo {author} {\bibfnamefont {M.}~\bibnamefont {Muller}}, \bibinfo
  {author} {\bibfnamefont {E.}~\bibnamefont {Rico}}, \bibinfo {author}
  {\bibfnamefont {P.}~\bibnamefont {Stebler}}, \bibinfo {author} {\bibfnamefont
  {U.-J.}\ \bibnamefont {Wiese}},\ and\ \bibinfo {author} {\bibfnamefont
  {P.}~\bibnamefont {Zoller}},\ }\href
  {https://doi.org/10.1103/PhysRevLett.109.175302} {\bibfield  {journal}
  {\bibinfo  {journal} {Phys. Rev. Lett.}\ }\textbf {\bibinfo {volume} {109}},\
  \bibinfo {pages} {175302} (\bibinfo {year} {2012})},\ \Eprint
  {https://arxiv.org/abs/1205.6366} {arXiv:1205.6366 [cond-mat.quant-gas]}
  \BibitemShut {NoStop}%
\bibitem [{\citenamefont {Martinez}\ \emph {et~al.}(2016)\citenamefont
  {Martinez}, \citenamefont {Muschik}, \citenamefont {Schindler}, \citenamefont
  {Nigg}, \citenamefont {Erhard}, \citenamefont {Heyl}, \citenamefont {Hauke},
  \citenamefont {Dalmonte}, \citenamefont {Monz}, \citenamefont {Zoller},\ and\
  \citenamefont {Blatt}}]{Martinez_2016}%
  \BibitemOpen
  \bibfield  {author} {\bibinfo {author} {\bibfnamefont {E.~A.}\ \bibnamefont
  {Martinez}}, \bibinfo {author} {\bibfnamefont {C.~A.}\ \bibnamefont
  {Muschik}}, \bibinfo {author} {\bibfnamefont {P.}~\bibnamefont {Schindler}},
  \bibinfo {author} {\bibfnamefont {D.}~\bibnamefont {Nigg}}, \bibinfo {author}
  {\bibfnamefont {A.}~\bibnamefont {Erhard}}, \bibinfo {author} {\bibfnamefont
  {M.}~\bibnamefont {Heyl}}, \bibinfo {author} {\bibfnamefont {P.}~\bibnamefont
  {Hauke}}, \bibinfo {author} {\bibfnamefont {M.}~\bibnamefont {Dalmonte}},
  \bibinfo {author} {\bibfnamefont {T.}~\bibnamefont {Monz}}, \bibinfo {author}
  {\bibfnamefont {P.}~\bibnamefont {Zoller}},\ and\ \bibinfo {author}
  {\bibfnamefont {R.}~\bibnamefont {Blatt}},\ }\href
  {https://doi.org/10.1038/nature18318} {\bibfield  {journal} {\bibinfo
  {journal} {Nature}\ }\textbf {\bibinfo {volume} {534}},\ \bibinfo {pages}
  {516} (\bibinfo {year} {2016})}\BibitemShut {NoStop}%
\bibitem [{\citenamefont {Muschik}\ \emph {et~al.}(2017)\citenamefont
  {Muschik}, \citenamefont {Heyl}, \citenamefont {Martinez}, \citenamefont
  {Monz}, \citenamefont {Schindler}, \citenamefont {Vogell}, \citenamefont
  {Dalmonte}, \citenamefont {Hauke}, \citenamefont {Blatt},\ and\ \citenamefont
  {Zoller}}]{Muschik:2016tws}%
  \BibitemOpen
  \bibfield  {author} {\bibinfo {author} {\bibfnamefont {C.}~\bibnamefont
  {Muschik}}, \bibinfo {author} {\bibfnamefont {M.}~\bibnamefont {Heyl}},
  \bibinfo {author} {\bibfnamefont {E.}~\bibnamefont {Martinez}}, \bibinfo
  {author} {\bibfnamefont {T.}~\bibnamefont {Monz}}, \bibinfo {author}
  {\bibfnamefont {P.}~\bibnamefont {Schindler}}, \bibinfo {author}
  {\bibfnamefont {B.}~\bibnamefont {Vogell}}, \bibinfo {author} {\bibfnamefont
  {M.}~\bibnamefont {Dalmonte}}, \bibinfo {author} {\bibfnamefont
  {P.}~\bibnamefont {Hauke}}, \bibinfo {author} {\bibfnamefont
  {R.}~\bibnamefont {Blatt}},\ and\ \bibinfo {author} {\bibfnamefont
  {P.}~\bibnamefont {Zoller}},\ }\href
  {https://doi.org/10.1088/1367-2630/aa89ab} {\bibfield  {journal} {\bibinfo
  {journal} {New J. Phys.}\ }\textbf {\bibinfo {volume} {19}},\ \bibinfo
  {pages} {103020} (\bibinfo {year} {2017})},\ \Eprint
  {https://arxiv.org/abs/1612.08653} {arXiv:1612.08653 [quant-ph]} \BibitemShut
  {NoStop}%
\bibitem [{\citenamefont {Kokail}\ \emph {et~al.}(2019)\citenamefont {Kokail},
  \citenamefont {Maier}, \citenamefont {van Bijnen}, \citenamefont {Brydges},
  \citenamefont {Joshi}, \citenamefont {Jurcevic}, \citenamefont {Muschik},
  \citenamefont {Silvi}, \citenamefont {Blatt}, \citenamefont {Roos},\ and\
  \citenamefont {Zoller}}]{Kokail_2019}%
  \BibitemOpen
  \bibfield  {author} {\bibinfo {author} {\bibfnamefont {C.}~\bibnamefont
  {Kokail}}, \bibinfo {author} {\bibfnamefont {C.}~\bibnamefont {Maier}},
  \bibinfo {author} {\bibfnamefont {R.}~\bibnamefont {van Bijnen}}, \bibinfo
  {author} {\bibfnamefont {T.}~\bibnamefont {Brydges}}, \bibinfo {author}
  {\bibfnamefont {M.~K.}\ \bibnamefont {Joshi}}, \bibinfo {author}
  {\bibfnamefont {P.}~\bibnamefont {Jurcevic}}, \bibinfo {author}
  {\bibfnamefont {C.~A.}\ \bibnamefont {Muschik}}, \bibinfo {author}
  {\bibfnamefont {P.}~\bibnamefont {Silvi}}, \bibinfo {author} {\bibfnamefont
  {R.}~\bibnamefont {Blatt}}, \bibinfo {author} {\bibfnamefont {C.~F.}\
  \bibnamefont {Roos}},\ and\ \bibinfo {author} {\bibfnamefont
  {P.}~\bibnamefont {Zoller}},\ }\href
  {https://doi.org/10.1038/s41586-019-1177-4} {\bibfield  {journal} {\bibinfo
  {journal} {Nature}\ }\textbf {\bibinfo {volume} {569}},\ \bibinfo {pages}
  {355} (\bibinfo {year} {2019})},\ \Eprint {https://arxiv.org/abs/1810.03421}
  {arXiv:1810.03421 [quant-ph]} \BibitemShut {NoStop}%
\bibitem [{\citenamefont {Surace}\ \emph {et~al.}(2020)\citenamefont {Surace},
  \citenamefont {Mazza}, \citenamefont {Giudici}, \citenamefont {Lerose},
  \citenamefont {Gambassi},\ and\ \citenamefont {Dalmonte}}]{Surace_2020}%
  \BibitemOpen
  \bibfield  {author} {\bibinfo {author} {\bibfnamefont {F.~M.}\ \bibnamefont
  {Surace}}, \bibinfo {author} {\bibfnamefont {P.~P.}\ \bibnamefont {Mazza}},
  \bibinfo {author} {\bibfnamefont {G.}~\bibnamefont {Giudici}}, \bibinfo
  {author} {\bibfnamefont {A.}~\bibnamefont {Lerose}}, \bibinfo {author}
  {\bibfnamefont {A.}~\bibnamefont {Gambassi}},\ and\ \bibinfo {author}
  {\bibfnamefont {M.}~\bibnamefont {Dalmonte}},\ }\href
  {https://doi.org/10.1103/physrevx.10.021041} {\bibfield  {journal} {\bibinfo
  {journal} {Phys. Rev. X}\ }\textbf {\bibinfo {volume} {10}},\ \bibinfo
  {pages} {021041} (\bibinfo {year} {2020})},\ \Eprint
  {https://arxiv.org/abs/1902.09551} {arXiv:1902.09551 [cond-mat.quant-gas]}
  \BibitemShut {NoStop}%
\bibitem [{\citenamefont {Jordan}\ and\ \citenamefont
  {Wigner}(1928)}]{Jordan:1928wi}%
  \BibitemOpen
  \bibfield  {author} {\bibinfo {author} {\bibfnamefont {P.}~\bibnamefont
  {Jordan}}\ and\ \bibinfo {author} {\bibfnamefont {E.}~\bibnamefont
  {Wigner}},\ }\href {https://doi.org/10.1007/BF01331938} {\bibfield  {journal}
  {\bibinfo  {journal} {Z. Phys.}\ }\textbf {\bibinfo {volume} {47}},\ \bibinfo
  {pages} {631} (\bibinfo {year} {1928})}\BibitemShut {NoStop}%
\bibitem [{\citenamefont {Farhi}\ \emph {et~al.}(2000)\citenamefont {Farhi},
  \citenamefont {Goldstone}, \citenamefont {Gutmann},\ and\ \citenamefont
  {Sipser}}]{Farhi2000}%
  \BibitemOpen
  \bibfield  {author} {\bibinfo {author} {\bibfnamefont {E.}~\bibnamefont
  {Farhi}}, \bibinfo {author} {\bibfnamefont {J.}~\bibnamefont {Goldstone}},
  \bibinfo {author} {\bibfnamefont {S.}~\bibnamefont {Gutmann}},\ and\ \bibinfo
  {author} {\bibfnamefont {M.}~\bibnamefont {Sipser}},\ }\href@noop {} {\
  (\bibinfo {year} {2000})},\ \Eprint {https://arxiv.org/abs/quant-ph/0001106}
  {arXiv:quant-ph/0001106 [quant-ph]} \BibitemShut {NoStop}%
\bibitem [{\citenamefont {Farhi}\ \emph {et~al.}(2001)\citenamefont {Farhi},
  \citenamefont {Goldstone}, \citenamefont {Gutmann}, \citenamefont {Lapan},
  \citenamefont {Lundgren},\ and\ \citenamefont {Preda}}]{Farhi2001-mi}%
  \BibitemOpen
  \bibfield  {author} {\bibinfo {author} {\bibfnamefont {E.}~\bibnamefont
  {Farhi}}, \bibinfo {author} {\bibfnamefont {J.}~\bibnamefont {Goldstone}},
  \bibinfo {author} {\bibfnamefont {S.}~\bibnamefont {Gutmann}}, \bibinfo
  {author} {\bibfnamefont {J.}~\bibnamefont {Lapan}}, \bibinfo {author}
  {\bibfnamefont {A.}~\bibnamefont {Lundgren}},\ and\ \bibinfo {author}
  {\bibfnamefont {D.}~\bibnamefont {Preda}},\ }\href@noop {} {\bibfield
  {journal} {\bibinfo  {journal} {Science}\ }\textbf {\bibinfo {volume}
  {292}},\ \bibinfo {pages} {472} (\bibinfo {year} {2001})},\ \Eprint
  {https://arxiv.org/abs/quant-ph/0104129} {arXiv:quant-ph/0104129 [quant-ph]}
  \BibitemShut {NoStop}%
\bibitem [{\citenamefont {Peruzzo}\ \emph {et~al.}(2014)\citenamefont
  {Peruzzo}, \citenamefont {McClean}, \citenamefont {Shadbolt}, \citenamefont
  {Yung}, \citenamefont {Zhou}, \citenamefont {Love}, \citenamefont
  {Aspuru-Guzik},\ and\ \citenamefont {O'Brien}}]{Peruzzo2014-hk}%
  \BibitemOpen
  \bibfield  {author} {\bibinfo {author} {\bibfnamefont {A.}~\bibnamefont
  {Peruzzo}}, \bibinfo {author} {\bibfnamefont {J.}~\bibnamefont {McClean}},
  \bibinfo {author} {\bibfnamefont {P.}~\bibnamefont {Shadbolt}}, \bibinfo
  {author} {\bibfnamefont {M.~H.}\ \bibnamefont {Yung}}, \bibinfo {author}
  {\bibfnamefont {X.~Q.}\ \bibnamefont {Zhou}}, \bibinfo {author}
  {\bibfnamefont {P.~J.}\ \bibnamefont {Love}}, \bibinfo {author}
  {\bibfnamefont {A.}~\bibnamefont {Aspuru-Guzik}},\ and\ \bibinfo {author}
  {\bibfnamefont {J.~L.}\ \bibnamefont {O'Brien}},\ }\href@noop {} {\bibfield
  {journal} {\bibinfo  {journal} {Nat. Commun.}\ }\textbf {\bibinfo {volume}
  {5}},\ \bibinfo {pages} {4213} (\bibinfo {year} {2014})},\ \Eprint
  {https://arxiv.org/abs/1304.3061} {arXiv:1304.3061 [quant-ph]} \BibitemShut
  {NoStop}%
\bibitem [{\citenamefont {Farhi}\ \emph {et~al.}()\citenamefont {Farhi},
  \citenamefont {Goldstone},\ and\ \citenamefont {Gutmann}}]{farhi2014quantum}%
  \BibitemOpen
  \bibfield  {author} {\bibinfo {author} {\bibfnamefont {E.}~\bibnamefont
  {Farhi}}, \bibinfo {author} {\bibfnamefont {J.}~\bibnamefont {Goldstone}},\
  and\ \bibinfo {author} {\bibfnamefont {S.}~\bibnamefont {Gutmann}},\
  }\href@noop {} {\bibinfo {title} {A quantum approximate optimization
  algorithm}},\ \Eprint {https://arxiv.org/abs/1411.4028} {arXiv:1411.4028
  [quant-ph]} \BibitemShut {NoStop}%
\bibitem [{\citenamefont {Wecker}\ \emph {et~al.}(2015)\citenamefont {Wecker},
  \citenamefont {Hastings},\ and\ \citenamefont {Troyer}}]{Wecker_2015}%
  \BibitemOpen
  \bibfield  {author} {\bibinfo {author} {\bibfnamefont {D.}~\bibnamefont
  {Wecker}}, \bibinfo {author} {\bibfnamefont {M.~B.}\ \bibnamefont
  {Hastings}},\ and\ \bibinfo {author} {\bibfnamefont {M.}~\bibnamefont
  {Troyer}},\ }\href {http://dx.doi.org/10.1103/PhysRevA.92.042303} {\bibfield
  {journal} {\bibinfo  {journal} {Phys. Rev. A}\ }\textbf {\bibinfo {volume}
  {92}},\ \bibinfo {pages} {042303} (\bibinfo {year} {2015})},\ \Eprint
  {https://arxiv.org/abs/1507.08969} {arXiv:1507.08969 [quant-ph]} \BibitemShut
  {NoStop}%
\bibitem [{\citenamefont {Clemente}\ \emph {et~al.}(2020)\citenamefont
  {Clemente} \emph {et~al.}}]{Clemente:2020lpr}%
  \BibitemOpen
  \bibfield  {author} {\bibinfo {author} {\bibfnamefont {G.}~\bibnamefont
  {Clemente}} \emph {et~al.} (\bibinfo {collaboration} {QuBiPF
  Collaboration}),\ }\href {https://doi.org/10.1103/PhysRevD.101.074510}
  {\bibfield  {journal} {\bibinfo  {journal} {Phys. Rev. D}\ }\textbf {\bibinfo
  {volume} {101}},\ \bibinfo {pages} {074510} (\bibinfo {year} {2020})},\
  \Eprint {https://arxiv.org/abs/2001.05328} {arXiv:2001.05328 [hep-lat]}
  \BibitemShut {NoStop}%
\end{thebibliography}%

\end{document}